\documentclass[%
 reprint, 
amsmath, amssymb,
aip,
]{revtex4-2}

\usepackage{graphicx}
\usepackage{dcolumn}
\usepackage{bm}
\usepackage{comment}
\usepackage{float} 
\usepackage{amsmath} 
\usepackage{amssymb}
\usepackage{lipsum}
\usepackage{graphicx}

\usepackage[caption=false]{subfig}

\usepackage[utf8]{inputenc}
\usepackage[T1]{fontenc}
\usepackage{mathptmx}

\begin{document}

\preprint{AIP/123-QED}

\title{Emission limited logarithmic and power law transients in pump-probe spectroscopy of perovskites
}

\author{Pradeep R. Nair}

\affiliation{Department of Electrical Engineering, Indian Institute of Technology Bombay, Mumbai, India}

\date{\today}

\begin{abstract}
Optical pump-probe techniques like absorption spectroscopy and microwave conductivity are widely used to characterize the carrier dynamics in perovskites for optoelectronic applications. In contrast to the prevalent assumption of exponentials, here we predict the possibility of trap emission limited logarithmic and power-law transients. These predictions are validated by detailed numerical simulations and well supported by several experimental reports from recent literature.  Interestingly, these findings indicate the need to revisit the existing schemes which rely on simplified rate equations and exponential decays to estimate the recombination parameters from pump-probe spectroscopy. Accordingly, we suggest appropriate methodologies to back extract parameters related to trap distribution from such non-exponential transients. Indeed, the insights shared in this manuscript could fundamentally impact the usage and interpretation of transient spectroscopy for emerging materials for optoelectronic applications. 
\end{abstract}

\maketitle

\section{Introduction}
\begin {figure} [h]
  \centering
    \includegraphics[width=0.45\textwidth]{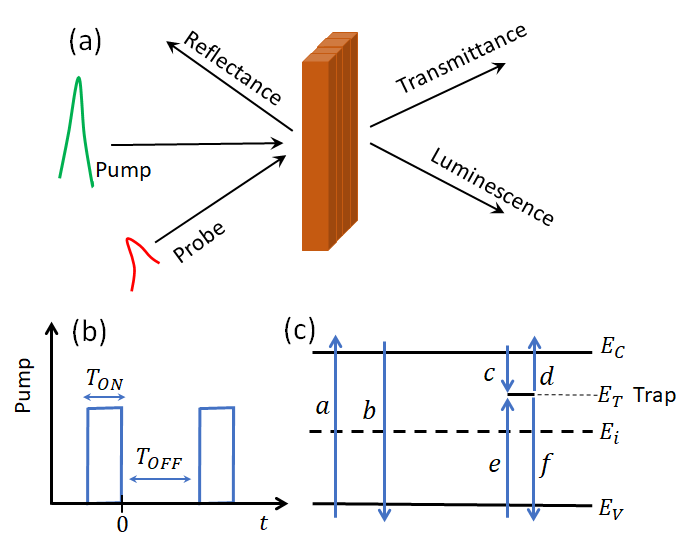}
    \caption{\textit{Pump-probe spectroscopy to characterize recombination processes. Part (a) shows a schematic in which the sample is optically excited by repeated pulses of pump signal (shown in part (b)). The carrier dynamics could be sensed through a probe signal and modulations in associated features like absorbance, reflectance, conductance, etc. Part (c) indicates the various electronic process in the semiconductor including photo-excitation, radiative recombination and trap assisted recombination. Although not explicitly shown, Auger recombination 
 is considered in later sections.}}
\label{schematic}
\end{figure}

Dynamics of carrier recombination and trapping is central to the operation of a wide variety of semiconductor devices including perovskite based solar cells\cite{stranks_review_recombination,park_review,motti_review} and light emitting diodes\cite{Fakharuddin2022}. Accordingly, diverse techniques are used to characterize such processes and associated parameters \cite{wang_review}. Among them, contact less exploration of materials is often crucial to understand the fundamental  electronic processes and further optimization of fabrication processes \cite{OPTP_TRMC_review,OPTP_review_chargedynamics}. In general, such characterization techniques involve pulsed mode of optical excitation (see Fig. \ref{schematic}a,b). During the ON period, carriers are generated due to the optical excitation. In the absence of excitation (i.e., OFF period), the carrier densities return, asymptotically, to its equilibrium state through various recombination mechanisms (Fig. \ref{schematic}c). Information regarding the recombination processes can be obtained with the help of a probe signal whose modulations vary linearly with carrier density\cite{OPTP_TRMC_review}. Examples of such pump-probe techniques are transient absorption\cite{TAS_review} ($TA$), diffuse reflectance \cite{diffuse_reflectance} ($DR$), microwave conductivity \cite{TRMC_ACSPhtonics} ($MC$), etc. On the other hand, Photoluminiscence ($PL$) varies as the product of carrier densities and does not involve any probe signal.\vspace{0.3em}\\ 
Theoretical models and numerical simulations are very essential in interpreting such experimental spectroscopy transients and further estimation of various parameters associated with recombination processes. The corresponding rate equations are readily available in well known references related to semiconductor device physics\cite{PierretADF}. Various phenomena detailed in Fig. \ref{schematic}c like radiative recombination, carrier capture by empty traps, emission from filled traps, trap assisted recombination, Auger recombination, etc. need to to be accounted in such a theoretical formalism. Further, the same formalism should consider not just single level traps but complex trap distributions like band tail states, trapping centers, etc. in a pulsed mode of operation - which can be computationally intensive. In fact, most of the recent modeling efforts on perovskites rely on over-simplified models which are ill-equipped to handle the influences of diverse defect/trap distributions and the associated trapping-emission limited regime with non-exponential transients (detailed arguments in this regard are provided in Sec. IIB).\vspace{0.3em}\\
In this context, here we predict, for the first time, that the pump-probe spectroscopy transients could vary logarithmically with time. Similar analysis also anticipates power-law transients as well. Indeed, several recent experimental data\cite{savinije_acsami_2021,savenije_acsel2020,herz_Bidoped,herz_ees2014,snaith_ees2022,pollock_jpcc,leijtens_trappingVoc} compare well with our predictions - even though the logarithmic or power-law trends in transients were not explicitly identified and analyzed in these literature reports. Indeed, such logarithmic or power-law transients are beyond the ambit of simple rate equations traditionally used to analyze  pump-probe spectroscopy transients. We further identify schemes to back extract information regarding trap distribution from non-exponential transients.\vspace{0.3em}\\
Below, we first detail the predictive models for emission limited regime which lead to non-exponential transients. The same is then validated through numerical simulations and compared against experimental results from literature.
 \section{Traps and carrier dynamics}
The schematic shown in Fig. \ref{schematic}a,b represents a typical pump-probe characterization scheme. Various electronic processes associated with a single trap level  is illustrated in part (c) (multiple levels or trap distribution will be considered later). As mentioned before, transients associated with many pump-probe spectroscopy schemes vary linearly with the  free carrier density\cite{OPTP_TRMC_review}. So, central to the analysis of such optical pump-probe characterization is the ability to predict the time dependent variation in carrier densities. In addition, such transients are expected to be dominated by the majority carrier density.\vspace{0.3em} \\
\textbf{A. Detailed model}: For the system shown in Fig. \ref{schematic}c with single level of traps, the photo-generated electrons could be either in the conduction band or at the trap states while the holes are in the valence band. In addition, the presence of charged defects and/or doping could also contribute to carrier density. Accordingly, charge neutrality indicates that  
  \begin{equation}
p=n+n_T\pm N_{d}
\label{eq:neutral}
 \end{equation}
where  $n$ and $p$ denote free electrons and holes, respectively, and $n_T$ denotes the density of  filled trap states. The term $N_{d}$ denotes the density of dopants (unintentional or otherwise) where $+$ and $-$ signs represent negatively and positively charged species, respectively.\vspace{0.3em} \\ 
In accordance with the various capture-emission processes depicted in Fig. \ref{schematic}c, the rate equations\cite{PierretADF} for $n$, $p$, and $n_T$ are 
 \begin{subequations}
\begin{align}
  \frac{\partial n}{\partial t}&= G -k_2(np-n_i^2) - C_{nT} + E_{nT} \\ 
  \frac{\partial p}{\partial t}&= G - k_2(np-n_i^2) -C_{pT}+E_{pT} \\
  \frac{\partial n_T}{\partial t}&=C_{nT}-E_{nT}-C_{pT}+E_{pT}
  \end{align}    
 \label{eq:gen_rate}
 \end{subequations}
 where $G$ denotes the photo-generation rate (process $a$ of Fig. \ref{schematic}c) and $k_2$ is the coefficient of radiative recombination (process $b$ of Fig. \ref{schematic}c). The parameter $C_{nT}$  denotes the rate of electron capture at empty traps (i.e., process $c$ of Fig. \ref{schematic}c) while $E_{nT}$ denotes the rate of electron emission from filled traps (i.e., process $d$ of Fig. \ref{schematic}c). Similar definition holds with respect to holes for $C_{pT}$ and $E_{pT}$. The rates under non-equilibrium conditions\cite{PierretADF} are given as 
 \begin{subequations}
\begin{align}
C_{nT}&=c_n(N_T-n_T)n\\
E_{nT}&=c_nn_Tn_1\\
C_{pT}&=c_p n_Tp\\
E_{pT}&=c_p(N_T-n_T)p_1
\end{align}
\label{eq:CE}
\end{subequations}
  where $c_n$ and $c_p$ denote the capture coefficients of electrons and holes by empty and filled traps, respectively.  $N_T$  denotes the density of total  traps (a fraction of which could be filled, $n_T$). The parameters $n_1$ and $p_1$ contain information regarding the energetic location of the traps. Application of detailed balance theory\cite{PierretADF} indicate that  $n_1=n_ie^{(E_{T'}-E_i)/kT}$ and $p_1=n_ie^{-(E_{T'}-E_i)/kT}$, where $n_i$ is the intrinsic carrier concentration, $E_{T'}$ is the effective trap level, $E_i$ is the intrinsic level, and $kT$ is the thermal energy (see Fig. \ref{schematic}c for the energy levels).\vspace{0.3em}\\ 
The above description of the various capture/emission processes for single level traps can be easily extended for any distribution of traps. While a chain of events dominated by  process $c$ and $e$ (see Fig. \ref{schematic}c) results in electron-hole recombination, dominance of processes $c$ and $d$ leads to trapping-emission - and not recombination.  Based on the above, the electron and hole trapping times are defined as $\tau_n=(c_nN_T)^{-1}$ and $\tau_p=(c_pN_T)^{-1}$, respectively.  The emission time from a filled trap depends on the trap depth from the conduction band. A quick inspection of eq. \ref{eq:gen_rate}c indicates that the emission time ($\tau_E$) from filled traps is given as 
  \begin{equation}
\tau_E=\tau_0 \times e^{E'/kT} 
     \label{eq:tauE}
 \end{equation}
where $\tau_0^{-1}=c_nN_C$, $E'=E_C-E_T'$, and $N_C$ is the conduction band effective density of states. Eq. \ref{eq:tauE} indicates that emission time scale exponentially with the trap depth from conduction band - i.e., deep traps emit much later than shallow traps. Later, we will see that this property has significant role in the emergence of non-exponential features in optical pump-probe transients. \vspace{0.3em}\\
The dominant role of a trap depends on the relative magnitudes of the rates described in eq. \ref{eq:CE}. For typical recombination centers, we have $C_{nT} \approx C_{pT}$ while the other two rates are negligible. However, all defects are not recombination centers - some could be just trapping centers. Such trapping centers could involve only in trapping and emission of carriers (i.e., processes $c$ and $d$ in Fig. \ref{schematic}b). The same can be accounted for electron trapping centers through the condition $E_{nT} >> C_{pT}$ (similar conditions apply for hole trapping centers as well). Hence, with appropriate parameters, the formalism described by eqs. \ref{eq:gen_rate}-\ref{eq:CE} can address different scenarios ranging from recombination to trapping-emission limited kinetics. As mentioned before, the description for a system with single trap level can be easily extended to more complex distributions. In addition, the influence of Auger recombination can be incorporated in eq. \ref{eq:gen_rate} through additional terms (discussed in Sec. VC).  Further, the same formalism can be used to  explore pulsed mode of operations as well ( see Fig. \ref{schematic}b) with an appropriate value for $G$ during the ON period (with $G=0$ during the OFF period).\vspace{0.3em}\\ 
\textbf{B. Simplified models used in perovskite literature}: Closed form analytical solutions for carrier transients (eqs. \ref{eq:gen_rate}-\ref{eq:CE}) are non-existent for the general case where all processes significant. Further, the analysis becomes more complex in the presence of trap distributions (instead of single level traps considered in eq. \ref{eq:gen_rate}). Hence, not surprisingly, several approximations of eq. \ref{eq:gen_rate} are widely used to explore carrier transients and hence interpret different optical pump-probe spectroscopy measurements. For example,  under quasi-steady state assumption for $n_T$ (i.e., $\frac{\partial n_T}{\partial t} \approx 0 $), the rate equation for $n$ is given as  $\frac{\partial n}{\partial t} = -k_{1}n-k_2np$, where $1/k_1$ is the effective time constant of trap-assisted recombination process. Under the limiting cases of dominant trap assisted recombination, the solutions are the well known exponential decays\cite{simon_lowK_lifetime} (i.e., $n(t) \propto e^{-k_1t}$). However, radiative recombination is not negligible in several cases. For example, transients in doped materials under low level injection conditions are exponential decays given as $n(t) \propto e^{-(k_1+k_2N_A)t}$, where $N_A$ is the effective doping density. \vspace{0.3em}\\
Another widely used approach to analyze transient spectroscopy has the following assumptions:  undoped material with $n=p$, $\tau_n=\tau_p$, and $\frac{\partial n_T}{\partial t} \approx 0$. The corresponding rate equation\cite{snaith_ees2022,herz_trapping_analysis} and general solution are given as   
  \begin{subequations}
\begin{align}
  \frac{\partial n}{\partial t} & = - k_{1}n-k_2n^2\\
n(t)=p(t)& = \frac{k_1}{-k_2+(k_2+k_1/n(0))e^{k_1t}}
  \end{align}   
   \label{eq:solkbb}
 \end{subequations}
where $n(0)$ is the carrier density at $t=0$. For $k_1 >> k_2$, eq. (\ref{eq:solkbb}) indicates that $n(t)$ decays exponentially with time. In the other limit $k_2 > >k_1$, the $n(t)$ varies as $1/t$. Typically, the values of $k_1$ and $k_2$ are back extracted by curve-fitting various forms of eq. \ref{eq:solkbb}.\vspace{0.3em}\\
A major drawback of the above approach is the inherent assumption that traps/defects act as recombination centers - which need not be universally correct. A few recent articles\cite{savinije_acsami_2021,savenije_acsel2020,pastor_PLdecay} indeed considered additional single level trapping centers (i.e., along with eq. \ref{eq:solkbb}a). However, closed form analytical solutions are not easily available for such cases and, as a consequence, parameter extraction is achieved through various forms of curve fitting. Further, the influence of trap distributions are yet to be explicitly considered. Indeed, in the absence of such analytical solutions, a-priori assumption of exponential or bi-exponential transients leads to inaccurate parameter estimation and the key insights otherwise directly evident in the experimental trends might remain unidentified. 
\begin {figure*} [ht!]
  \centering
    \includegraphics[width=0.9\textwidth]{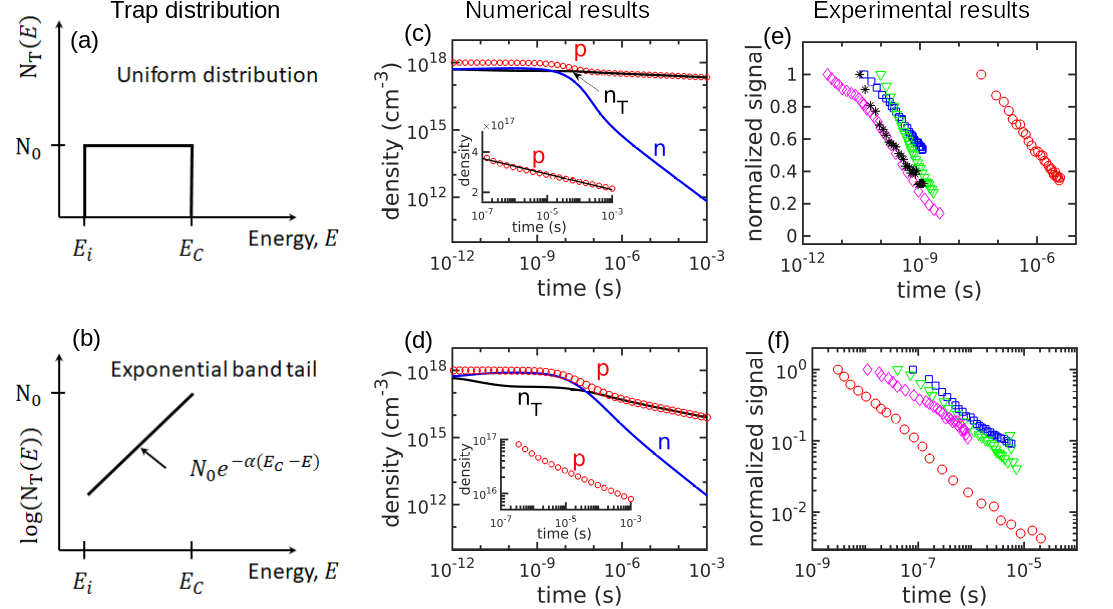}
    \caption{\textit{ Theoretical predictions and experimental validations on logarithmic and power-law transients in pump-probe spectroscopy. Left panel shows the trap distributions under consideration. Middle panel shows the corresponding results from numerical solutions of eqs. \ref{eq:gen_rate}-\ref{eq:CE}. Right panel shows experimental results from literature. Parts (a) and (b) illustrate uniform distribution and exponential band tail trapping centers, respectively. Simulation results corresponding to uniform trap distribution is shown in part (c). The inset of the same (with Y axis in linear scheme and X axis in log scale) clearly show that majority carriers ($p$) vary logarithmically with time. Part (d) shows simulation results corresponding to  exponential band tail states. Here the long term transients for majority carriers vary as a power-law with time (i.e., linear variation in a log-log scheme). Experimental results from literature indicate logarithmic trends (part (e)) and power-law transients (part (f)). The pump-probe spectroscopy techniques used to obtain the experimental data shared in parts (e) and (f) are either transient absorption ($TA$) or microwave conductivity ($MC$). The experimental data in part (e) are from the following sources: red (MC, Fig. 1b of ref.\cite{savinije_acsami_2021}), blue (TA, Fig. 2b of ref. \cite{snaith_ees2022}), green (TA, Fig. 2 of ref. \cite{herz_ees2014}), black (TA, Fig. 4b of ref. \cite{herz_Bidoped}) and magenta (TA, Fig. S7c of ref. \cite{pollock_jpcc}). The experimental data in part (f) are from the following sources: red (TA, Fig. 3b of ref. \cite{deAngelis_powerlaw}), blue (MC, abstract graphic of ref. \cite{savenije_acsel2020}), green (TA, Fig. 2b of ref. \cite{leijtens_trappingVoc}), and magenta (MC, Fig. 4b of ref. \cite{savinije_acsami_2021}). Theoretical predictions on logarithmic and power-law transients are well supported by both numerical simulations and experimental data from literature. }}
\label{Fig1}
\end{figure*}
\section{Emission limited Non-exponential transients}
In view of the above-listed various shortcomings of the state-of-art analysis schemes for pump-probe spectroscopy, here we develop an analytical model to address the influence of trapping centers on the  carrier transients. We first consider uniform trap distribution (Fig. \ref{Fig1}a) and then exponential band tail states ((Fig. \ref{Fig1}b)). The same is represented by eqs. \ref{eq:gen_rate}-\ref{eq:CE} with the condition $C_{pT}=E_{pT}=0$. The influence of additional mid-gap recombination centers and Auger recombination will be considered later.\vspace{0.3em}\\
\textbf{A. Logarithmic transients}: Consider a system with uniform density of electron trapping centers denoted by $N_T(E)=N_0$, as shown in Fig. \ref{Fig1}a. At energy $E$, the density of filled traps is given as $n_T(E)$. During the ON condition, a significant fraction of the traps will be filled as denoted by $n_T=\int_{E_i}^{E_C} n_T(E) dE$. Note the convention - parameters without explicit mention of $E$ denote integrated quantities. During the decay transients, two phenomena attempt to modulate $n$: (a) radiative recombination process, and (b) trapping-emission between conduction band and the traps. Although equal amounts of electrons and holes are created by the pump signal, some trap states will be filled by electrons. Hence, as per eq. \ref{eq:neutral}, we have $p > n$ as the initial conditions for the decay transients. Accordingly, the resultant long-term spectroscopy signal will be dominated by transients in $p$  and hence that of $n_T$ as well (as per eq. \ref{eq:neutral} and since $p>>n$).\vspace{0.3em}\\
The time dependence for $n_T$ can be obtained as follows: Assume that all traps up to a certain energy are filled at $t=0$. During the OFF transients, free electron density in the conduction band, $n$, reduces due to recombination with free holes in the valence band. As time progresses, electron emission from filled traps attempt to repopulate the conduction band. Hence, at any time $t$, one can assume that all traps with emission time $\tau_E< t$ are near empty. Rather, only those traps with corresponding $\tau_E>t$ will remain filled at time $t$. Hence $n_T(t)=N_0(E_{T'}(t)-E_V)$ where $E_{T'}(t)$ denotes the trap level with emission time $t_E=t$. As $t_E$ is given by eq. \ref{eq:tauE}, we have 
  \begin{equation}
p(t) \approx n_T(t) \propto N_0 kT \times ln(t)
     \label{eq:log}
 \end{equation}
 which predicts that the majority carrier density varies logarithmically with time. Since the transient spectroscopy signal is proportional to majority carrier density, the same is also expected to vary logarithmically with time.\vspace{0.3em}\\ 
\textbf{B. Power-law transients}: In the presence of band-tail states (see Fig. \ref{Fig1}b), the transients could show significantly different trends. Here, the distribution is  given as $N_T(E')=N_0 e^{-\alpha E'}$ where $\alpha^{-1}$ denotes the Urbach energy and $E'=E_C-E_{T'}$ denotes the trap depth from the conduction band. The behavior of long-term transients can be obtained through an analysis similar to the previous case. Here also we assume that at time $t$, all traps with $\tau_E < t$ are near empty while the rest are completely filled. Hence, at  $t=\tau_E$,  we have $n_T(t) = \int_{E'(t)}^{0.5E_g} N_T(E') dE'$, where $E_g$ is the band gap.  Again, using eq. \ref{eq:tauE} for $\tau_E$ and the condition $p>>n$, we find 
  \begin{equation}
p(t)\approx n_T(t) \propto \frac{N_0}{\alpha} \times {\left(\frac{t}{\tau_0}\right)}^{-\alpha kT}
 \label{eq:power}
 \end{equation}
 Equation \ref{eq:power} predicts that the majority carrier transients could vary as power-law with time exponents uniquely defined by the Urbach energy. Note that this time exponent ($-\alpha kT$) is distinctly different from the predicted exponent of $-1$ for the radiative recombination limited scenario (see eq. \ref{eq:solkbb}b). We further remark that the above  prediction of power-law transients are broadly consistent with with earlier work on phosphorescence \cite{randall_wilkins}, transient diffusion reflectance\cite{Nandal_nature} and absorption\cite{Vikas_design} in photoanodes for water-splitting, and disordered semiconductors\cite{seki_powerlaw}. Notably, these earlier reports on power law exponents were arrived through more complex analysis compared to the simple derivation provided in this manuscript. However, logarithmic transients, predicted by eq. \ref{eq:log}, are yet to be reported in literature. \vspace{0.3em}\\
  The transients in electron density in both cases is closely connected to the quasi-steady trapping-emission between conduction band and the relevant trap level and the recombination mechanisms. Under such conditions, trend for $n(t)$ can be obtained as follows: The net rate of recombination in all cases is  $\partial p /\partial t$, and not $\partial n /\partial t$. This is due to the fact that a reduction in hole density, for the scenario under consideration, happens only through  recombination while changes in  $n$ could be due to trapping-emission and recombination. Hence, for the case of uniform density of traps with $p$ varying as $ln(t)$, the net recombination is expected to vary as $t^{-1}$. With dominant radiative recombination, and given the fact that $p$ varies slowly with $t$, we expect that the $n(t)$ varies as $t^{-1}$. Similar arguments can be extended for exponential band tail states as well. As before, the net recombination is nothing but $\partial p/ \partial t$. With $p \propto t^{-\alpha kT}$, we find that the net recombination should vary as $t^{-(\alpha kT+1)}$. Accordingly, under dominant radiative recombination, we expect that $n$ varies as $t^{-1}$ for exponential band tail states. \vspace{0.3em}\\
 The above discussion indicates that under dominant radiative recombination, regardless of whether the trap distribution is uniform or exponential band tail states, the electron density is expected to vary as $t^{-1}$. Similar arguments for the case with additional trap assisted recombination via mid-gap states indicate that the electron transients would vary still vary as $t^{-1}$ for uniform density of trapping centers. On the other hand, for exponential band tail states, it can be shown that the $n(t)$ varies as $t^{-(\alpha kT+1)}$.  
\section{Validation of theoretical predictions}
\textbf{A. Numerical simulations}: Having theoretically established the possibilities of trap distribution dependent logarithmic and power-law transients, here we validate the same first through detailed numerical simulations and then with experimental results from literature. Here we consider two scenarios (see Fig. \ref{Fig1}a,b for schematic representation of trap distributions) - (a) uniform density of traps ($N_0=6\times 10^{17}cm^{-3}eV^{-1}$), and (b) exponential band tail states ($\alpha=1/4kT$ and $N_0=4.5\times 10^{18}cm^{-3}eV^{-1}$). The integrated density of traps ($N_T$) is given as $N_T=\int_{E_i}^{E_C} N_T(E) dE$. Accordingly, for uniform distribution of traps  with $E_g=1.6eV$, we have  $N_T=4.8 \times 10^{17}cm^{-3}$. For exponential band tail states, similar analysis yields $N_T=4.5 \times 10^{17}cm^{-3}$. This indicates that the  cases under consideration differ only in their energetic distribution and not in terms of the integrated density of traps - which could allow us to clearly associate the salient features of the transients with the energetic distribution of traps.\vspace{0.3em}\\
For each of the above mentioned cases, the decay transients are explored through numerical solutions of eqs. \ref{eq:gen_rate}-\ref{eq:CE} with $G=0$ and $k_2=10^{-10}cm^3s^{-1}$. The initial conditions used are (i.e., at at $t=0$) : (i) all traps are filled (i.e., $n_T=N_T$), and (ii) $p=10^{18}cm^{-3}$ with $n=p-N_T$. Each distribution is considered as a combination of several closely spaced trap levels. Eqs. \ref{eq:gen_rate}-\ref{eq:CE}, appropriately modified to account for such multiple levels (each with $c_p=0$ and $c_n=10^{-8}cm^3s$), are numerically solved to obtain the carrier transients.\vspace{0.3em}\\
Figures \ref{Fig1}c,d shows the numerical simulation results for the two trap distributions under consideration. The long-term trends indicate that the majority carrier density ($p$) varies logarithmically with time 
 for uniform density of trapping centers (Fig. \ref{Fig1}c and its inset) and as power law for exponential band tail states (Fig. \ref{Fig1}d) - thus validating the analytical predictions. Further, the transients in $n$ for both cases vary as $t^{-1}$ - in accordance with model predictions.\vspace{0.3em}\\
 The trends during the initial times are also amenable to easy interpretation. The initial condition is that all traps are filled with $n < p$. With $p \approx 10^{18}cm^{-3}$ and $k_2 = 10^{-10}cm^3s^{-1}$, the radiative lifetime, $(k_2p)^{-1}$, for minority electrons is $10^{-8}s$. Hence, the most dominant mechanism at initial times (i.e., $t < 10^{-8}s$) is the emission or de-trapping of electrons from filled traps.  As a result, $n$ increases initially. This is more prominent in Fig. \ref{Fig1}d with exponential band tail states as more electrons occupy trap states with lower emission times. For $t \sim 10^{-8}s$, radiative recombination becomes significant. However, as $n<p$, further reduction due to radiative recombination is more pronounced in $n$ than in $p$.  Once $p$ becomes similar in magnitude to $n_T$, further asymptotic decrease in $p$ follows either eq. \ref{eq:log} or eq. \ref{eq:power}.\vspace{0.3em} \\ 
Interestingly, the numerical validation of analytical predictions is much more quantitative than just the dominant time dependence. For example, the slope of $p$ vs. $ln(t)$ obtained from simulations (see inset of Fig. \ref{Fig1}c) is $-1.53 \times 10^{16}$, which as per eq. \ref{eq:log}, should be $N_0 \times kT$. The back extracted value for $N_0$ from the slope is $6 \times 10^{17} cm^{-3}eV^{-1}$ which is the same as the value used in simulations. Similarly, the power law transients in Fig. \ref{Fig1}c also compares, quantitatively, very well with eq. \ref{eq:power}. The power law exponent obtained from simulations ($-0.258$) is closely predicted by the theory ($-\alpha kT$, with $\alpha=1/4kT$). In addition, the pre-factor obtained from numerical solutions ($ \sim 4.3 \times 10^{14}$) is also well anticipated by the analytical expressions ($N_0 \alpha^{-1} \tau_0^{-\alpha kT}$). Such features can be used to back-extract key parameters associated to trap distributions from experimental data. \vspace{0.3em}\\
\textbf{B. Experimental results from literature}: Our predictions on logarithmic and power law transients are well supported by recent experimental results as well. For example, Fig. \ref{Fig1}e clearly shows logarithmic time dependence in results from multiple labs on diverse materials and spectroscopy measurements. The logarithmic time dependence is conclusive as the experimental results shared in Fig. \ref{Fig1}e vary linearly in a $y$ vs. $ln(x)$ scheme for about 2 orders of magnitude variation in time. In addition, although not included in Fig. \ref{Fig1}e, results another reference\cite{huang_log} also show clear logarithmic trends. Interestingly, although the normalized trends are plotted, the slopes of these experimentally observed logarithmic transients are very similar. Hence eq. \ref{eq:log} indicates that these samples are dominated by uniform trap distributions with similar magnitude.\vspace{0.3em}\\
Our predictions on power-law transients are also well supported by experimental results from literature - as shown in  Fig. \ref{Fig1}f. As before these are research results from independent labs under diverse conditions. Here, the back extracted exponent from power-law transients was around $-0.64$ for data from references\cite{deAngelis_powerlaw,savenije_acsel2020,leijtens_trappingVoc} and the corresponding Urbach energy (using eq. \ref{eq:power}) is $40 meV$. The experimental data from ref.\cite{savinije_acsami_2021} has $-0.46$ as the power law exponent with a correponding Urbach energy of $56meV$. These estimates, although not impossible, are somewhat larger than other reports in literature\cite{wolf_urbach} for halide perovskites and hence would require further analysis.\vspace{0.3em}\\
 The experimental validation of theoretical predictions is all the more remarkable due to the fact that only one data set used in Figs. \ref{Fig1}e,f was plotted in a form where the logarithmic (i.e., $y$ vs. $ln(x)$) or power-law trends could be easily identified (i.e., $ln(y)$ vs. $ln(x)$, ref. \cite{deAngelis_powerlaw}). The rest of the data were reported either as $y$ vs. $x$ or as $ln(y)$ vs. $x$ - schemes typically associated with exponential variations. In the absence of theoretical models, eqs. \ref{eq:log}-\ref{eq:power}, the analysis available in literature is not detailed enough to identify the trap distributions involved - single level vs. uniform vs. Urbach band tail states and their influence on the eventual device performance. 
\section{Discussions} 
\begin {figure} [h]
  \centering
\includegraphics[width=0.45\textwidth]{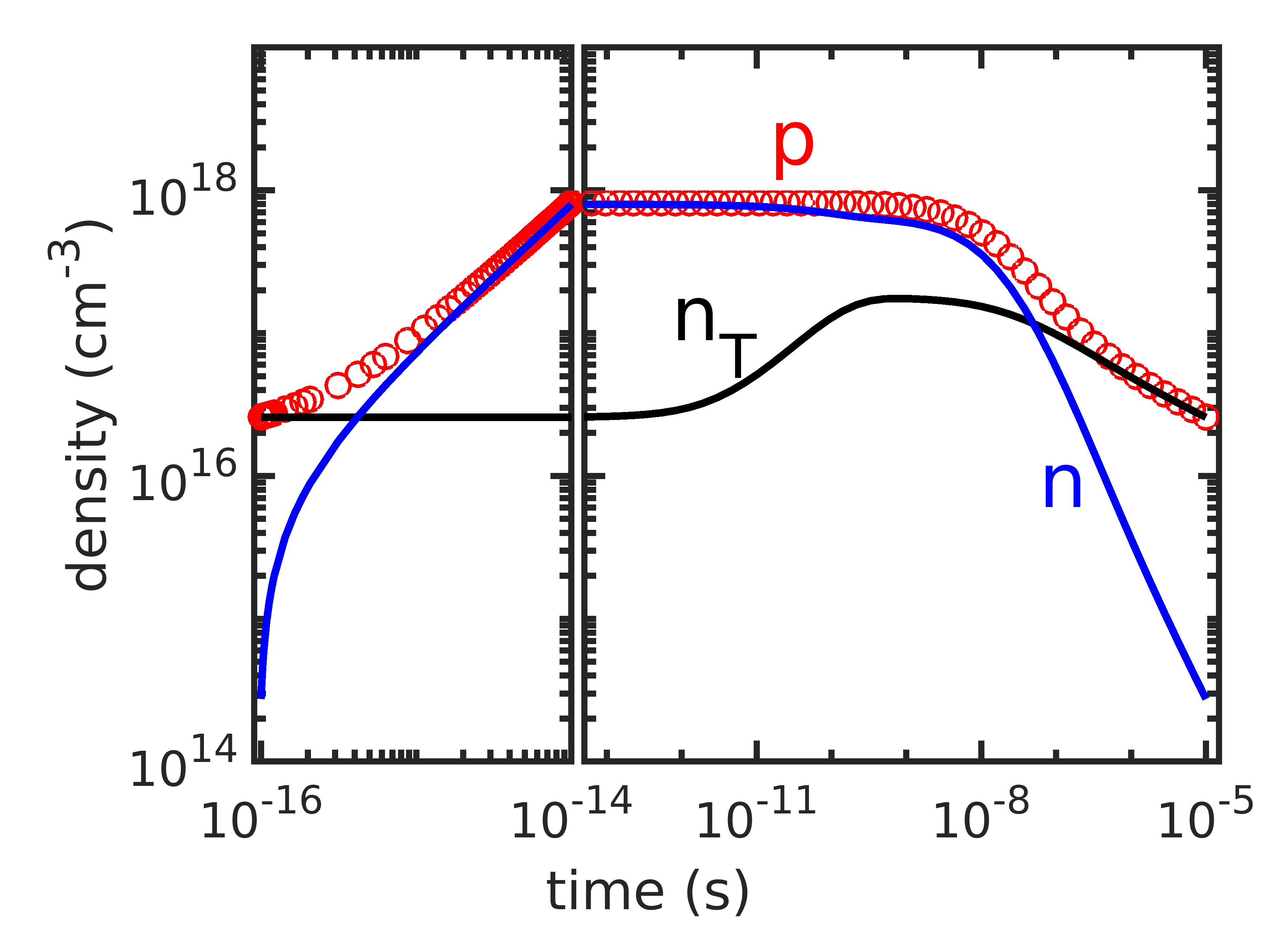}
    \caption{\textit{Influence of pulsed operations on carrier transients. The  transients shown correspond to ON (left panel) and OFF (right panel) stages with pump repetition frequency of 10MHz. The OFF transients in carrier density are similar to the results shared in Fig. \ref{Fig1}c. However, the transients in $n_T$ are significantly different during the initial stages.}}
\label{cycle_fig}
\end{figure}

\textbf{A. Pulsed Operations}: The formalism described by eqs. \ref{eq:gen_rate}-\ref{eq:CE} can be used to address pulsed mode of operations as well. In a pulsed mode of operation, the carrier densities increase due to photo-generation during the ON phase which are followed by decay transients during the OFF phase (see Fig. \ref{schematic}b). It is evident that with appropriate $G$ in each phase, eqs. \ref{eq:gen_rate}-\ref{eq:CE} can be solved over multiple time periods (or cycles) to obtain the required transient response. Importantly, the results obtained after a large number of such cycles is independent of the initial conditions used for the first cycle.\vspace{0.3em}\\
Simulation results of such a pulsed mode of operation with a net carrier generation of $G \times T_{ON}=10^{18}cm^{-3}$ per pump pulse at a repetition frequency of $10MHz$ is shown in Fig. \ref{cycle_fig} (band tail states with $\alpha=1/4kT$). Here, $T_{ON}=10fs$ is the duration in which the pump signal is ON. During the ON phase, the carrier densities increase almost linearly with time (see left panel of Fig. \ref{cycle_fig}). However, trap-filling usually lags behind. As a consequence, significant amount of trap filling happens over the initial duration of the OFF phase (i.e., $n_T$ increases even with $G=0$, see the right panel of Fig. \ref{cycle_fig}) with a corresponding reduction in $n$. Further reduction in $n$ occurs due to radiative recombination. Later, emission from traps dominate and power-law transients emerge for both $p$ and $n_T$. In comparison with the results shared in Fig. \ref{Fig1}c, the major difference in pulsed-mode of operations is in the transients of $n_T$. However, the long-term transients for carrier densities retain the same characteristics, as predicted by eqs. \ref{eq:log}-\ref{eq:power}.\\

\begin {figure} [h]
  \centering
    \includegraphics[width=0.43\textwidth]{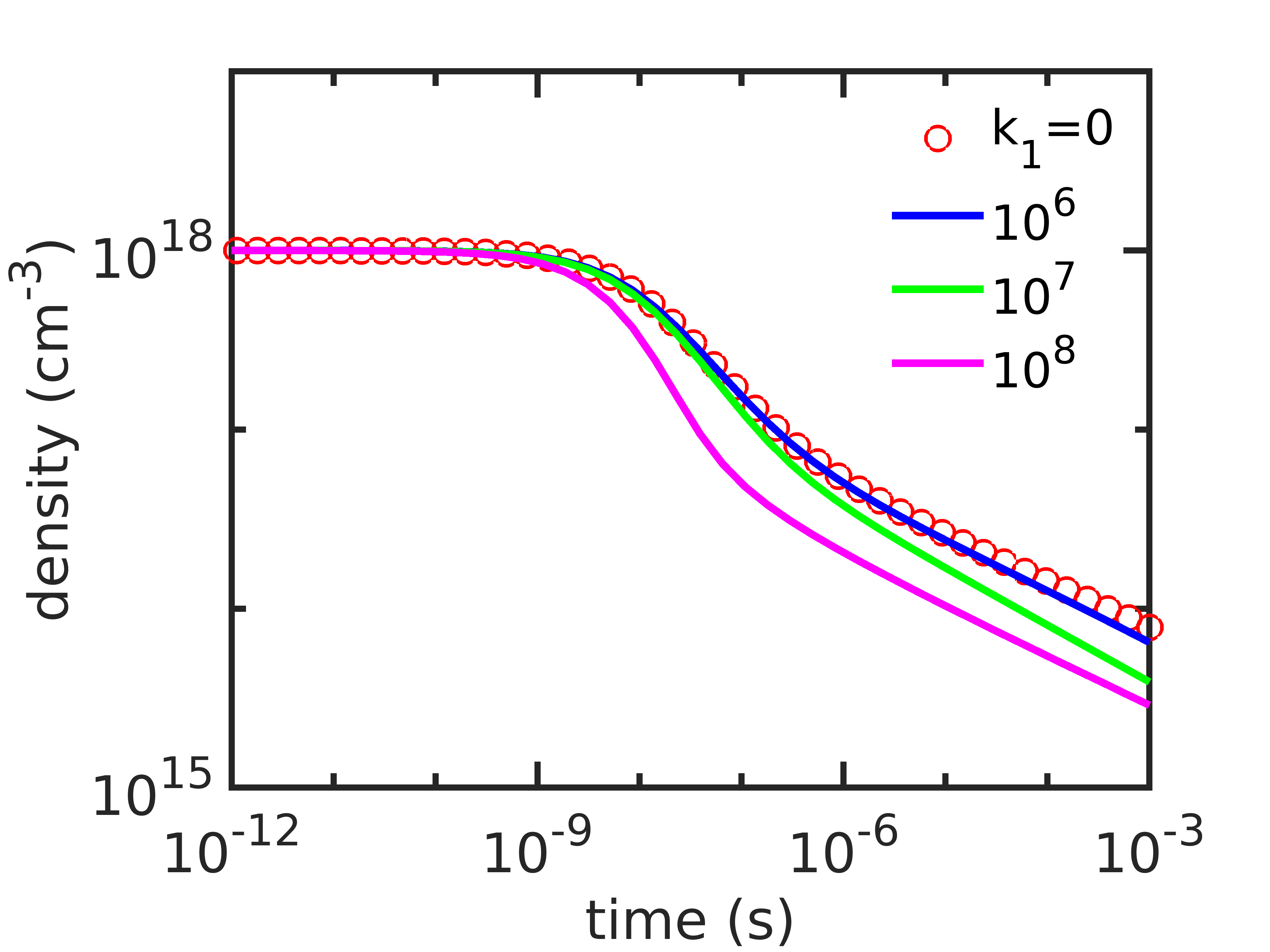}
    \caption{\textit{Simulation results on the combined influence of mid-gap recombination centers and exponential band tail trapping centers on the carrier transients. Legends indicate the value of $k_1$ (units: $s^{-1}$) used in respective simulations.}}
\label{lifetime}
\end{figure}

\textbf{B. Combined influence of trapping and recombination centers}: The influence of trapping centers on the decay transients were elucidated so far. However, in addition to the such trapping centers, mid-gap recombination centers could also be present in most perovskite materials. The corresponding lifetimes are of the order of several hundreds of nanoseconds. For a system with single level trapping centers and mid-gap recombination centers, the updated set of equations which describe the transients are as follows:
 \begin{equation}
\begin{aligned}
  \frac{\partial n}{\partial t}&= G -k_1n-k_2(np-n_i^2) - C_{nT} + E_{nT} \\ 
  \frac{\partial p}{\partial t}&= G -k_1n- k_2(np-n_i^2)\\
    \frac{\partial n_T}{\partial t}&= C_{nT} - E_{nT}  
 \end{aligned}    
 \label{eq:recomb}
 \end{equation}
 As discussed before, the above system can be easily updated for various trap distributions like uniform and band tail states. We performed simulations for different values of $k_1^{-1}$ ranging from $10\mu s$ to $10ns$ along with additional exponential band tail trapping centers. The results for the same, shared in Fig. \ref{lifetime}, indicate that the power law nature of long-term transients, remain unchanged even in the presence of additional mid-gap states (except with minor changes in the power law exponents). Interestingly, mono-molecular lifetimes larger than $1 \mu s$ have negligible influence on the transients at all (compare between the data sets corresponding to $k_1=0$ and $k_1=10^6s^{-1}$ in Fig. \ref{lifetime}). Further, even with monomolecular lifetimes of the order of tens of nanoseconds, the long-term transients are not exponentials.  Instead, the nature of long-term transients are governed by the emission process from filled traps.  Similar trends hold good for uniform density of traps as well. 

\textbf{C. Estimation of recombination parameters}: So far, we have conclusively shown, through numerical simulations and experimental results from literature, that trap emission manifests as logarithmic or power law transients in pump-probe spectroscopy. Back extraction of key parameters is an expected outcome from characterization experiments. Here, the power law transients allow easy estimation of Urbach energy. A possible scheme for characterization of the trap distribution using eqs. \ref{eq:log}-\ref{eq:power} has been explicitly mentioned along with the discussion of simulation results (see Sec. IVA). Specifically,  the only unknown for logarithmic transients is the uniform trap density $N_0$. This could be extracted from the slope of un-normalized transients with explicit knowledge of $c_1$, the proportionality constant between measured transients and carrier density. Similarly, the unknowns in power-law transients are $\alpha$ and $N_0$. The first could be obtained from the exponent of power-law transients while the second could be obtained from the intercepts, as detailed in Sec. IVA.\vspace{0.3em}\\
Effective carrier lifetime dictates\cite{motti_review,stranks_review_recombination,nair_jpcl2014,nair_APL2015,nandal2017,abhimanyu_JAP2021} the performance of many optoelectronic devices like solar cells and LEDs. The same is influenced by the parameters $k_1$ and $k_2$. In view of logarithmic and power-law transients, and given the relative lack of sensitivity of carrier transients to $k_1$ (see eqs. \ref{eq:log}-\ref{eq:power} and Fig. \ref{lifetime}), is it possible to reliably estimate $k_1$ and $k_2$ from pump-probe spectroscopy transients alone? The feasibility depends on the  the carrier densities at the beginning of OFF transients - specifically, the values of $n$ and $p$. As evident in the simulation results (see Fig. \ref{Fig1}c,d), the early part of the transients is more influenced by the recombination phenomena while the later parts could be emission limited. Radiative recombination dominates trap assisted recombination for $n>>k_1/k_2$. For typical values like $k_1 = 10^6s^{-1}$ and $k_2=10^{-10}cm^3s^{-1}$, such transition happens for $n \approx 10^{16}cm^{-3}$. With large $n$ as the initial condition (i.e., $n >> 10^{16}cm^{-3}$), it is possible to estimate $k_2$ from the early part of transients before any logarithmic or power-law (with exponent other than $-1$) trends become evident (see the discussions, Sec IVA).\vspace{0.3em}\\
The results in Fig. \ref{lifetime} indicate that the influence of $k_1$ is negligible during the later transients (unless $k_1$ is very large). This indicates that $k_1$ should be estimated from the initial parts of OFF transients where radiative recombination is negligible. Similar concerns are valid to characterize Auger recombination which vary as $k_3n^3$, where $k_3$ is the coefficient of Auger recombination. It can be shown that Auger recombination becomes appreciable only at $n >> k_2/k_3$. With reported value\cite{herz_ees2014} of $k_3 \approx 10^{-29}cm^6s^{-1}$ and $k_2 \approx 10^{-10}cm^3s^{-1}$, the corresponding $n$ evaluates to $\approx 10^{19}cm^{-3}$. Generation of such large carrier densities would require extremely large pump intensities, as shown below.\vspace{0.3em}\\
In pulsed mode of  experiments (see Fig. \ref{schematic}b), usually  the photon flux (i.e., net number of photons incident, $F$) or the energy density associated with the pulse ($E_P$) is specified. It is evident that not all of the incident photons could be absorbed by the active material (due to any reflectance/transmittance loss). A quick estimate for the photo-generation rate $G$ during the ON phase is given as  
   \begin{equation}
G= \frac{E_P }{T_{ON}}\frac{\lambda}{hc}\frac{f}{t_{abs}}
     \label{eq:G_exp}
 \end{equation}
where $\lambda$ is the wavelength of the incident photons, $f$ denotes the fraction of incident photons absorbed, $T_{ON}$ is the ON time of the pulse, $h$ is the Planck's constant, $c$ is the velocity of light, and $t_{abs}$ is the thickness of the active material. As discussed in Sec. VA, the initial carrier concentrations for the OFF transients in a pulsed mode of operation is $GT_{ON}$. For $GT_{ON}=10^{19}cm^{-3}$, with $\lambda=400nm$, $t_{abs}=250nm$, and $f=1$ (i.e., an unlikely best case scenario), the energy density of the pump pulse turns out to be $E_P \approx 125 \mu J/cm^2$ - which is much larger than the values used in experiments\cite{snaith_ees2022,savinije_acsami_2021} (which are of the order of tens of $\mu J/cm^2$). Usually $f<1$, and hence the corresponding practical estimates for $E_P$ could be much larger than $125 \mu J/cm^2$.\vspace{0.3em}\\
The above discussions clearly indicate that the recombination parameters could be unambiguously back extracted only through a set of spectroscopy measurements with a broad range of variation in the initial conditions for $n$ and $p$ (i.e., by varying the pump magnitude). Later part of the transients could be emission limited and contains information regarding trapping centers and not recombination centers. Further, curve-fitting different forms of eq. \ref{eq:solkbb} to logarithmic or power-law transients results in inaccurate estimates for recombination parameters. Indeed, given its slow varying nature, it is not surprising to obtain large lifetimes if an exponential fit is applied to logarithmic or power-law transients. Importantly, such  back extracted values for $k_1$ might result in inaccurate estimates for the eventual device performance. Hence, simple exponential fits to experimental data of limited range is of limited relevance towards device optimization. As such the dominant functional dependence of experimental data should be first ascertained before any parameters are back-extracted using curve-fitting exponentials or simplified rate equations (eq. \ref{eq:solkbb}).
\section{Conclusions}
To summarize, here we elucidated the influence of trapping centers on the transient pump-probe spectroscopy for perovskites. Contrary to the conventional analysis schemes which assumes exponential transients, here we predicted and numerically validated the emergence of trap distribution dependent logarithmic and power-law transients. The same are well supported by several experiment results from multiple labs reported in literature using diverse techniques and materials. We also identified appropriate schemes for back extraction of relevant parameters from such non-exponential transients. These insights are of fundamental and immense relevance towards contact-less material characterization for opto-electronic applications. 
\begin{acknowledgments}
This project is funded in part by Science and Engineering Research Board (SERB, project code: CRG/2019/003163), Department of Science and Technology (DST), India. PRN acknowledges Visvesvaraya Young Faculty Fellowship and discussions with Tahir Patel, Dr. Nithin Chatterji, and Prof. K. L. Narasimhan.
\end{acknowledgments}

\section*{References}

\bibliography{aipsamp}

\end{document}